\documentclass[runningheads]{llncs}

\usepackage[utf8]{inputenc}
\usepackage{color}
\usepackage{graphicx}
\usepackage{hyperref}
\usepackage{tabularx}
\usepackage{multirow}
\usepackage{booktabs}
\usepackage{footmisc}
\usepackage[normalem]{ulem}
\useunder{\uline}{\ul}{}
\usepackage{paralist}
\usepackage{array}
\usepackage[misc]{ifsym}

\newcolumntype{M}[1]{>{\centering\arraybackslash}m{#1}}

\makeatletter
\newcommand{\printfnsymbol}[1]{%
  \textsuperscript{\@fnsymbol{#1}}%
}
\makeatother

\begin{document}
\title{Deep Learning Models for Multilingual Hate Speech Detection\thanks{Accepted at ECML-PKDD 2020}}
%
%

\author{Sai Saketh Aluru\inst{1}\thanks{Equal Contribution}\and
Binny Mathew\inst{1}\printfnsymbol{2} \and
Punyajoy Saha\inst{1} \and \\
Animesh Mukherjee \inst{2} \\
Indian Institute of Technology Kharagpur, India
}




\authorrunning{Aluru et al.}

\institute{\email{\{saisakethaluru, binnymathew, punyajoys\}@iitkgp.ac.in} \and
\email{animeshm@cse.iitkgp.ac.in}}
%

%
%
\maketitle              
\setcounter{footnote}{0}
\begin{abstract}

Hate speech detection is a challenging problem with most of the datasets available in only one language: English. In this paper, we conduct a large scale analysis of multilingual hate speech in 9 languages from 16 different sources. We observe that in low resource setting, simple models such as LASER embedding with logistic regression performs the best, while in high resource setting BERT based models perform better. In case of zero-shot classification, languages such as Italian and Portuguese achieve good results. Our proposed framework could be used as an efficient solution for low-resource languages. These models could also act as good baselines for future multilingual hate speech detection tasks. We have made our code and experimental settings public\footnote{\url{https://github.com/punyajoy/DE-LIMIT}} for other researchers.

\end{abstract}

\section{Introduction}
Online social media has allowed dissemination of information at a faster rate than ever~\cite{mathew2019spread,mathew2019temporal}. This has allowed bad actors to use this for their nefarious purposes such as propaganda spreading, fake news, and \textit{hate speech}. Hate speech is defined as a ``direct and serious attack on any protected category of people based on their race, ethnicity, national origin, religion, sex, gender, sexual orientation, disability or disease''~\cite{elsherief2018hate}. Representative examples of hate speech are provided in Table~\ref{tab:examples}.

Hate speech is increasingly becoming a concerning issue in several countries. Crimes related to hate speech have been increasing in the recent times with some of them leading to severe incidents such as the genocide of the Rohingya community in Myanmar, the anti-Muslim mob violence in Sri Lanka, and the Pittsburg shooting. Frequent and repetitive exposure to hate speech has been shown to desensitize the individual to this form of speech and subsequently to lower evaluations of the victims and greater distancing, thus increasing outgroup prejudice~\cite{soral2018exposure}. The public expressions of hate speech has also been shown to affect the devaluation of minority members~\cite{greenberg1985effect}, the exclusion of minorities from the society~\cite{mullen2003ethnophaulisms}, and the discriminatory distribution of public resources~\cite{fasoli2015labelling}. 

While the research in hate speech detection has been growing rapidly, one of the current issues is that majority of the datasets are available in English language only. Thus, hate speech in other languages are not detected properly and this could be detrimental. This is a problem for companies like Facebook as well, which can detect hate speech in certain languages only (English, Spanish, and Mandarin)\footnote{\url{https://time.com/5739688/facebook-hate-speech-languages/}}. While there are few datasets~\cite{basile2019semeval,ousidhoum2019multilingual} in other language available, as we observe, they are relatively small in size.

In this paper, we perform the first large scale analysis of multilingual hate speech by analyzing the performance of deep learning models on 16 datasets from 9 different languages. We consider two different scenarios and discuss the classifier performance. In the first scenario (monolingual setting), we only consider the training and testing from the same language. We observe that in low resource scenario models using LASER embedding with Logistic regression perform the best, whereas in high resource scenario, BERT based models perform much better. We also observe that simple techniques such as translating to English and using BERT, achieves competitive results in several languages. In the second scenario (multilingual setting), we consider training data from all the other languages and test on one target language. Here, we observe that including data from other languages is quite effective especially when there is almost no training data available for the target language (aka zero shot). Finally, from the summary of the results that we obtain, we construct a catalogue indicating which model is effective for a particular language depending on the extent of the data available. We believe that this catalogue is one of the most important contributions of our work which can be readily referred to by future researchers working to advance the state-of-the-art in multilingual hate speech detection.

The rest of the paper is structured as follows. Section~\ref{sec:relatedworks} presents the related literature for hate speech classification. In section \ref{sec:dataset}, we present the datasets used for the analysis. Section ~\ref{sec:experiment} provides details about the models and experimental settings. In section \ref{sec:results}, we note the key results of our experiments. In section \ref{sec:discussion} we discuss the results and provide error analysis.

\begin{table}[!tb]
\centering
\scriptsize
\caption{Examples of hate speech.}
\label{tab:examples}
\begin{tabular}{p{10cm} c}
Text                                                                              & Hate Speech? \\ \hline
I f**king hate ni**ers!                                                           & Yes          \\
Jews are the worst people on earth and we should get rid of them.                    & Yes          \\
“6 million was not enough. next time ovens will be the least of your concerns \# sixmillionmore” & Yes \\
Mexicans are f**king great people!                                                & No          
\end{tabular}

\end{table}

\section{Related Works}\label{sec:relatedworks}
Hate speech lies in a \textit{complex nexus with freedom of expression, individual, group and minority rights, as well as concepts of dignity, liberty and equality}~\cite{gagliardone2015countering}. Computational approaches to tackle hate speech has recently gained a lot of interest. The earlier efforts to build hate speech classifiers used simple methods such as dictionary look up~\cite{guermazi2007using}, bag-of-words~\cite{burnap2016us}.
 Fortuna \textit{et al.}~\cite{fortuna2018survey} conducted a comprehensive survey on this subject.

With the availability of larger datasets, researchers started using complex models to improve the classifier performance. These include deep learning~\cite{Badjatiya:2017:DLH:3041021.3054223,zhang2018detecting} and graph embedding techniques~\cite{ribeiro2018characterizing} to detect hate speech in social media posts. Zhang \textit{et al.}~\cite{zhang2018detecting} used deep neural network, combining convolutional and gated recurrent networks to improve the results on 6 out of 7 datasets used. In this paper, we have used the same CNN-GRU model for one of our experimental settings (monolingual scenario).

Research into the multilingual aspect of hate speech is relatively new. Datasets for languages such as Arabic and French~\cite{ousidhoum2019multilingual}, Indonesian~\cite{ibrohim2019multi}, Italian~\cite{sanguinetti2018italian}, Polish~\cite{ptaszynski2019results}, Portuguese~\cite{fortuna2019hierarchically}, and Spanish~\cite{basile2019semeval} have been made available for research. To the best of our knowledge, very few works have tried to utilize these datasets to build multilingual classifiers. Huang \textit{et al.}~\cite{huang2020multilingual} used Twitter hate speech corpus from five languages and annotated them with demographic information. Using this new dataset they study the demographic bias in hate speech classification. Corazza \textit{et al.}~\cite{corazza2020multilingual} used three datasets from three languages (English, Italian, and German) to study the multilingual hate speech. The authors used models such as SVM, and Bi-LSTM to build hate speech detection models. Our work is different from these existing works as we perform the experiment on a much larger set of languages (9) using more datasets (16). Our work tries to utilize the existing hate speech resources to develop models that could be generalized for hate speech detection in other languages.

\section{Dataset description}\label{sec:dataset}

We looked into the datasets available for hate speech and found 16 publicly\footnote{Note that although Table~\ref{tab:dataset_stats} contains 19 entries, there are three occurrences of Ousidhoum \textit{et al.} \cite{ousidhoum2019multilingual} and two occurrences of Basile \textit{et al.}~\cite{basile2019semeval} for different languages.} available sources in 9 different languages\footnote{We relied on \url{hatespeechdata.com} for most of the datasets.}. One of the immediate issues, we observed was the mixing of several types of categories (offensive, profanity, abusive, insult etc). Although these categories are related to hate speech, they should not be considered as the same~\cite{davidson2017automated}. For this reason, we only use two labels: \textit{hate speech} and \textit{normal}, and discard other labels. Next, we explain the datasets in different languages. The overall dataset statistics are noted in Table~\ref{tab:dataset_stats}.

\noindent\textbf{Arabic:} We found two arabic datasets that were built for hate speech detection.
\begin{compactenum}
    \item [-] Mulki \textit{et al.} \cite{mulki2019hsab} : A Twitter dataset\footnote{\url{https://github.com/Hala-Mulki/L-HSAB-First-Arabic-Levantine-HateSpeech-Dataset}} for hate speech and abusive language. For our task, we ignored the abusive class and only considered the hate and normal class.
    \item [-] Ousidhoum \textit{et al.} \cite{ousidhoum2019multilingual}: A Twitter dataset\footnote{\url{https://github.com/HKUST-KnowComp/MLMA_hate_speech}\label{link:ousidhoum}} with multi-label annotations. We have only considered those datapoints which have either hate speech or normal in the annotation label.
\end{compactenum}

\begin{table}[!htb]
\centering
\caption{Dataset details}
\label{tab:dataset_stats}
\begin{tabular}{llllll}
Language                 & Dataset & Source        & Hate & Non-Hate & Total \\\hline
\multirow{2}{*}{Arabic} & Mulki \textit{et al.}~\cite{mulki2019hsab} &   Twitter & 468 & 3,652    & 4,120 \\
         & Ousidhoum \textit{et al.}~\cite{ousidhoum2019multilingual} &  Twitter &   755 &  915 & 1,670 \\\hline
\multirow{7}{*}{English} & Davidson \textit{et al.}~\cite{davidson2017automated}    & Twitter       & 1,430 & 4,163    & 5,593 \\
                         &  Gibert \textit{et al.}~\cite{de2018hate}     &    Stormfront  & 1,196 & 9,748     & 10,944 \\
                         & Waseem \textit{et al.}~\cite{waseem2016hateful} &  Twitter & 759  & 5,545     & 6,304  \\
                         & Basile \textit{et al.}~\cite{basile2019semeval} &   Twitter &  5,390 & 7,415     & 12,805 \\
                    &Ousidhoum \textit{et al.}~\cite{ousidhoum2019multilingual} &    Twitter &    1,278   &   661 &  1,939 \\
                    &Founta \textit{et al.}~\cite{founta2018large} & Twitter &  4,948& 53,790  &    58,738 \\\hline
\multirow{2}{*}{German} & Ross \textit{et al.}~\cite{ross2017measuring} &   Twitter   &   54  &  315 & 369 \\
         & Bretschneider \textit{et al.}~\cite{bretschneider2017detecting} & Facebook    & 625 &   5,161   &   5,786 \\\hline
\multirow{2}{*}{Indonesian} & Ibrohim \textit{et al.}~\cite{ibrohim2019multi} &    Twitter &   5,561   &  7,608   & 13,169 \\
         & Alfina \textit{et al.}~\cite{alfina2017hate} & Twitter & 260 &    453 &   713 \\\hline         
\multirow{2}{*}{Italian} & Sanguinetti \textit{et al.}~\cite{sanguinetti2018italian}&  Twitter & 231 &    1,329   &   1,560 \\
         &Bosco \textit{et al.}~\cite{bosco2018overview} &  Facebook \& Twitter &  3,355 & 4,645   &    8,000 \\\hline         
\multirow{1}{*}{Polish} &Ptaszynski \textit{et al.}~\cite{ptaszynski2019results} &   Twitter &   598    &  9,190   & 9,788 \\\hline   
\multirow{1}{*}{Portuguese} &Fortuna \textit{et al.}~\cite{fortuna2019hierarchically}& Twitter & 1,788    &    3,882   &   5,670 \\\hline
\multirow{2}{*}{Spanish} &Basile \textit{et al.}~\cite{basile2019semeval}&   Twitter &   2,228  &  3,137   & 5,365 \\
         &Pereira \textit{et al.}~\cite{pereira2019detecting} &   Twitter &   1,567   &  4,433   & 6,000 \\\hline
\multirow{1}{*}{French} &Ousidhoum \textit{et al.}~\cite{ousidhoum2019multilingual}&    Twitter &   399 &   821 &  1,220 \\\hline
Total & & &32,890 & 126,863 & 159,753\\\hline
\end{tabular}

, 
\end{table}
\noindent\textbf{English:} Majority of the hate speech datasets are available in English language. We select six such publicly available datasets.
\begin{compactenum}
    \item [-] Davidson \textit{et al.} \cite{davidson2017automated} provided a three class Twitter dataset\footnote{\url{https://github.com/t-davidson/hate-speech-and-offensive-language}}, the classes being hate speech, abusive speech, and normal. We have only considered the hate speech and normal class for our task.
    \item [-] Gibert \textit{et al.} \cite{de2018hate} provided a hate speech dataset\footnote{\url{https://github.com/aitor-garcia-p/hate-speech-dataset}} consisting sentences from Stormfront\footnote{\url{www.stormfront.org}}, a white supremacist forum. Each sentence is tagged as either hate or normal.
    \item [-] Waseem \textit{et al.} \cite{waseem2016hateful} provided a Twitter dataset\footnote{\url{https://github.com/zeerakw/hatespeech}} annotated into classes: sexism, racism, and neither. We considered the tweets tagged as sexism or racism as hate speech and neither class as normal.
    \item [-] Basile \textit{et al.} \cite{basile2019semeval} provided multilingual Twitter dataset\footnote{\url{https://github.com/msang/hateval}\label{link:basile}} for hate speech against immigrants and women. Each post is tagged as either hate speech or normal.
    \item [-] Ousidhoum \textit{et al.} \cite{ousidhoum2019multilingual} provided Twitter dataset\footref{link:ousidhoum} with multi-label annotations. We have only considered those datapoints which have either hate speech or normal in the annotation label.
    \item [-] Founta \textit{et al.} \cite{founta2018large} provided a large dataset\footnote{\url{https://github.com/ENCASEH2020/hatespeech-twitter}} of 100K annotations divided in four classes: hate speech, abusive, spam, and normal. For our task, we have only considered the datapoints marked as either hate or normal, and ignored the other classes.
\end{compactenum}

\noindent\textbf{German:} We select two datasets available in German language.

\begin{compactenum}
    \item [-] Ross \textit{et al.} \cite{ross2017measuring} provided a German hate speech dataset\footnote{\url{https://github.com/UCSM-DUE/IWG_hatespeech_public}} for the refugee crisis. Each tweet is tagged as hate speech or normal.
    \item [-] Bretschneider \textit{et al.} \cite{bretschneider2017detecting} provided a Facebook hate speech dataset\footnote{\url{http://www.ub-web.de/research/}} against foreigners and refugees.
\end{compactenum}

\noindent\textbf{Indonesian} We found two datasets for the Indonesian language.

\begin{compactenum}
    \item [-] Ibrohim \textit{et al.} \cite{ibrohim2019multi} provided an Indonesian multi-label hate speech and abusive dataset\footnote{\url{https://github.com/okkyibrohim/id-multi-label-hate-speech-and-abusive-language-detection}}. We only consider the hate speech label for our task and other labels are ignored.
    \item [-] Alfina \textit{et al.} \cite{alfina2017hate} provided an Indonesian hate speech dataset\footnote{\url{https://github.com/ialfina/id-hatespeech-detection}}. Each post is tagged as hateful or normal.
\end{compactenum}

\noindent\textbf{Italian} We found two datasets for the Italian language.
\begin{compactenum}
    \item [-] Sanguinetti \textit{et al.} \cite{sanguinetti2018italian} provided an Italian hate speech dataset\footnote{\url{https://github.com/msang/hate-speech-corpus}} against the minorities in Italy.
    \item [-] Bosco \textit{et al.} \cite{bosco2018overview} provided hate speech dataset\footnote{\url{https://github.com/msang/haspeede2018}} collected from Twitter and Facebook.
\end{compactenum}

\noindent\textbf{Polish} We found only one dataset for the Polish language
\begin{compactenum}
    \item [-] Ptaszynski \textit{et al.} \cite{ptaszynski2019results} provided a cyberbullying dataset\footnote{\url{http://poleval.pl/tasks/task6}} for the Polish language. We have only considered hate speech and normal class for our task.
\end{compactenum}

\noindent\textbf{Portuguese} We found one dataset for the Portuguese language

\begin{compactenum}
    \item [-] Fortuna \textit{et al.} \cite{fortuna2019hierarchically} developed a hierarchical hate speech dataset\footnote{\url{https://github.com/paulafortuna/Portuguese-Hate-Speech-Dataset}} for the Portuguese language. For our task, we have used the binary class of hate speech or normal.

\end{compactenum}

\noindent\textbf{Spanish} We found two dataset for the Spanish language.

\begin{compactenum}
    \item [-] Basile \textit{et al.} \cite{basile2019semeval} provided multilingual hate speech dataset\footref{link:basile} against immigrants and women.
    \item [-] Pereira \textit{et al.} \cite{pereira2019detecting} provided hate speech dataset\footnote{\url{https://zenodo.org/record/2592149}} for the Spanish language.
\end{compactenum}

\noindent\textbf{French}

\begin{compactenum}
    \item [-] Ousidhoum \textit{et al.} \cite{ousidhoum2019multilingual} provided Twitter dataset\footref{link:ousidhoum} with multi-label annotations. We have only considered those data points which have either hate speech or normal in the annotation label.
\end{compactenum}

\section{Experiments}\label{sec:experiment}

For each language, we combine all the datasets and perform stratified train/ validation/ test split in the ratio 70\%/10\%/20\%. For all the experiments, we use the same splits of train/val/test. Thus, the results are comparable across different models and settings. We report macro F1-score to measure the classifier performance. In case we select a subset of the dataset for the experiment, we repeated the subset selection with 5 different random sets and report the average performance. This would help to reduce the performance variation across different sets. In our experiments, the subsets are stratified samples of size $16,32,64,128,256$. 

\subsection{Embeddings}
In order to train models in multilingual setting, we need multilingual word/sentence embeddings. For sentences, LASER embeddings were used and for words MUSE embeddings were used.\\
\textbf{Laser embeddings:} LASER\footnote{\url{https://github.com/facebookresearch/LASER}} denotes Language-Agnostic SEntence Representations~\cite{artetxe2019massively}. Given an input sentence, LASER provides sentence embeddindgs which are obtained by applying max-pooling operation over the output of a BiLSTM  encoder. The system uses a single BiLSTM encoder with a shared BPE vocabulary for all languages.

\noindent\textbf{Muse embeddings:} MUSE\footnote{\url{https://github.com/facebookresearch/MUSE}} denotes Multilingual Unsupervised and Supervised Embeddings. Given an input word, MUSE gives as output the  corresponding word embedding~\cite{conneau2017word}. MUSE builds a bilingual dictionary between two languages without using any parallel corpora, by aligning monolingual word embedding spaces in an unsupervised way.

\subsection{Models}

\textbf{CNN-GRU (Zhang \textit{et al.} \cite{zhang2018detecting}}): This model initially maps each of the word in a sentence into a 300 dimensional vector using the pretrained Google News Corpus embeddings~\cite{mikolov2013efficient}. It also pads/clips the sentences to a maximum of 100 words. Then this $300 \times 100$ vector is passed through drop layer and finally to a 1-D convolution layer with 100 filters. Further, a maxpool layer reduces the dimension to $25  \times 100$ feature matrix. Now this is passed through a GRU layer and it outputs a $100 \times 100$ dimension matrix which is globally max-pooled to provide a $1 \times 100$ vector. This is further passed through a softmax layer to give us the final prediction.

\noindent\textbf{BERT:} BERT~\cite{devlin2018bert} stands for Bidirectional Encoder Representations from Transformers pretrained on data from english language. It is a stack of transformer encoder layers with multiple ``heads'', i.e. fully connected neural networks augmented with a self attention mechanism. For every input token in a sequence, each head computes key value and query vectors which are further used to create a weighted representation. The outputs of each head in the same layer are combined and run through a fully connected layer. Each layer is wrapped with a skip connection and a layer normalization is applied after it. In our model we set the token length to 128 for faster processing of the query\footnote{In the total data 0.17\% datapoints have more than 128 tokens when tokenized, thus justifying our choice.}.

\noindent\textbf{mBERT:} Multilingual BERT (mBERT~\footnote{https://tinyurl.com/yxh57v3a}) is a version of BERT that was trained on Wikipedia in 104 languages. Languages with a lot of data were sub-sampled and others were super sampled and the model was pretrained using the same method as BERT. mBERT generalizes across some scripts and can retrieve parallel sentences. mBERT is simply trained on a multilingual corpus with no language IDs, but it encodes language identities. We used mBERT to train hate speech detection model in different languages once again limiting to a maximum of 128 tokens for sentence representation. 

\noindent\textbf{Translation:} One simple way to utilize datasets in different languages is to rely on translation. Simple techniques of translation has shown to give good results in tasks such as sentiment analysis~\cite{singhal2016borrow}. We use Google Translate\footnote{\url{https://github.com/sergei4e/gtrans}} to convert all the datasets in different languages to English since translation to English from other languages typically have less errors in comparison to the other way round.

For our experiments we use the following four models:

\begin{enumerate}
    \item \textbf{MUSE + CNN-GRU:} For the given input sentence, we first obtain the corresponding MUSE embeddings which are then passed as input to the CNN-GRU model.
    \item \textbf{Translation + BERT:} The input sentence is first translated to the English language which are then provided as input to the BERT model.
    \item \textbf{LASER + LR:} For the given input sentence, we first obtain the corresponding LASER embeddings which are then passed as input to a Logistic Regression (LR) model.
    \item \textbf{mBert:} The input sentence is directly fed to the mBert model.
\end{enumerate}

\subsection{Hyperparameter optimization}
We use the validation set performance to select the best set of hyperparameters for the test set. The hyperparameters used in our experiments are as follows: batch size: $16$,
learning rate: $2e^{-5}, 3e^{-5}, 5e^{-5}$ and
epochs: $1,2,3,4,5$.

\if{0}
shown in Table~\ref{Tab:fine_tune_parameter}.

\begin{table}[htb]
\scriptsize
\centering
\caption{Fine-tuning hyperparameters.}
\label{Tab:fine_tune_parameter}
\begin{tabular}{ll}
\toprule
Hyperparameters & Values           \\
\midrule
Batch size    & $16$               \\
Learning Rate & $2e^{-5}, 3e^{-5}, 5e^{-5}$ \\
Epochs        & $1,2,3,4,5$       \\
\bottomrule
\end{tabular}
\end{table}
\fi

\section{Results}\label{sec:results}


\subsection{Monolingual scenario}
In this setting, we use the data from the same language for training, validation and testing. This scenario commonly occurs in the real world where monolingual dataset is used to build classifiers for a specific language.

\noindent\textbf{Observations:} Table~\ref{tab:samelanguage} reports the results of the monolingual scenario. As expected, we observe that with increasing training data, the classifier performance increases as well. However, the relative performance seem to vary depending on the language and the model. We make several observations. First, \textbf{LASER + LR} performs the best in low-resource settings (16,32,64,128,256) for all the languages. Second, we observe that \textbf{MUSE + CNN-GRU} performs the worst in almost all the cases. Third, \textbf{Translation + BERT} seems to achieve competitive performance for some of the languages such as German, Polish, Portuguese, and Spanish.
Overall we observe that there is no `one single recipe' for all languages; however, \textbf{Translation + BERT} seems to be an excellent compromise. We believe that improved translations in some languages can further improve the performance of this model. 

Although \textbf{LASER + LR} seems to be doing good in low resource setting, if enough data is available, we observe that BERT based models: \textbf{Translation + BERT} (English, German, Polish, and French) and  \textbf{mBERT} (Arabic, Indonesian, Italian, and Spanish) are doing much better. However, what is more interesting is that although BERT based models are known to be successful when a larger number of datapoints are available, even with 256 datapoints some of these models seem to come very close to \textbf{LASER + LR}; for instance, \textbf{Translation + BERT} (Spanish, French) and \textbf{mBERT} (Arabic, Indonesian, Italian).

\begin{table}[!tb]
\scriptsize
\centering
\caption{Monolingual scenario: the training, validation and testing data is used from the same language. Here, Full D represents the full training data. The \textbf{bold} figures represent the best scores and \underline{underline} represents the second best.}
\label{tab:samelanguage}
\resizebox{.90\textwidth}{!}{%
\begin{tabular}{lllllllll}
\toprule
\multirow{2}{*}{Language}   & \multirow{2}{*}{Model} & \multicolumn{6}{c}{Training Size}         \\ 
                            &                        & 16 & 32 & 64 & 128 & 256 & Full D \\ \midrule
\multirow{5}{*}{Arabic}     & MUSE + CNN-GRU        &0.4412    &0.4438    &0.4486     &0.4664    &0.5818       &0.7368        \\
                            & Translation + BERT    &0.4555    &0.4495    &\underline{0.5551}    &0.5448     &0.7017       &\underline{0.8115}        \\
                            & LASER + LR             &\textbf{0.5533}    &\textbf{0.6755}    &\textbf{0.7304}    &\textbf{0.7488}     &\textbf{0.7698}    &0.7920 \\
                            & mBert                  &\underline{0.4588}  &\underline{0.4533}  &0.4408  &\underline{0.6486}   &\underline{0.7295}   &\textbf{0.8320}        \\ \midrule
\multirow{5}{*}{English}    & MUSE + CNN-GRU        &\underline{0.4580}    &\underline{0.4594}    &\underline{0.4653}    &0.4646     &\underline{0.4813}     &0.6441        \\
                            &  BERT    &0.4071    &0.3925    &0.4260    &\underline{0.4720}     &0.4578     &\textbf{0.7143}        \\
                            & LASER + LR             &\textbf{0.4617}    &\textbf{0.4899}    & \textbf{0.5376}    & \textbf{0.5624}     &\textbf{0.5885}  &0.6526 \\
                            & mBert                  &0.1773    &0.3251    &0.4488    &0.4578     &0.4578     &\underline{0.7101}               \\ \midrule
\multirow{5}{*}{German}     & MUSE + CNN-GRU        &0.4708    &0.4708    &0.4708    &0.4708     &0.4762     &0.5756              \\
                            & Translation + BERT    &0.4812    &\underline{0.4758}    &\underline{0.4719}    &\underline{0.4729}     &0.4724     &\textbf{0.7662}               \\
                            & LASER + LR             &\underline{0.4974}    &\textbf{0.5201}    &\textbf{0.5465}    &\textbf{0.5925}     &\textbf{0.6488}     &\underline{0.6873}        \\
                            & mBert                  &\textbf{0.5037}    &0.4750    &0.4708    &0.4717     &\underline{0.5022}     &0.6517              \\ \midrule
\multirow{5}{*}{Indonesian} & MUSE + CNN-GRU        &0.4250    &0.4823    &0.5263    &0.5354     &0.5890     &0.7110               \\
                            & Translation + BERT    &0.4957    &0.5003    &0.5179    &0.5682     &0.6341     &0.7670               \\
                            & LASER + LR             &\textbf{0.5226}    &\textbf{0.5376}    &\textbf{0.5882}    &\textbf{0.6259}     &\textbf{0.6890}     &\underline{0.7872}\\
                            & mBert                  & \underline{0.5106} & \underline{0.5219} &\underline{0.5414}  & \underline{0.6016}  & \underline{0.6530}  & \textbf{0.8119}              \\ \midrule
\multirow{5}{*}{Italian}    & MUSE + CNN-GRU        &0.4055    &0.4476    &0.4461    &0.5206     &0.5965     &0.7349               \\
                            & Translation + BERT    &0.5006    &\underline{0.5943}    &\underline{0.6215}    &\underline{0.6678}     &0.6919     &0.7922               \\
                            & LASER + LR             &\underline{0.5688}    &\textbf{0.6210}    &\textbf{0.6843}    &\textbf{0.7175}     &\textbf{0.7347}     &\underline{0.7996}        \\
                            & mBert                  & \textbf{0.5774} & 0.4567 & 0.5834 & 0.6664  & \underline{0.7026}  & \textbf{0.8260}              \\ \midrule
\multirow{5}{*}{Polish}     & MUSE + CNN-GRU        &\underline{0.4842}    &0.4842    &0.4841    &\underline{0.4842}     &\underline{0.5180}     &0.6337               \\
                            & Translation + BERT    &\underline{0.4842}    &\underline{0.4853}    &\underline{0.4842}    &\underline{0.4842}     &0.5066     &\textbf{0.7161}               \\
                            & LASER + LR             &\textbf{0.4889}    &\textbf{0.4879}    &\textbf{0.5360}    &\textbf{0.5739}     &\textbf{0.6172}     &0.6439\\
                            & mBert                  &0.4829    &0.4847    &\underline{0.4842}    &\underline{0.4842}     &0.4842     &\underline{0.7069}               \\ \midrule
\multirow{5}{*}{Portuguese}  & MUSE + CNN-GRU        &0.4480    &0.3807    &0.4184    &0.4228     &0.4562     &0.6100               \\
                            & Translation + BERT    &0.4532    &\underline{0.4893}    &\underline{0.4712}    &0.5102     &\underline{0.5994}     &\underline{0.6935}               \\
                            & LASER + LR             &\textbf{0.5194} &\textbf{0.5536}  &\textbf{0.6070}  &\textbf{0.6210}  &\textbf{0.6412}     &\textbf{0.6941}\\
                            & mBert                  &\underline{0.5154}  & 0.4245 & 0.4148 & \underline{0.5493}  & 0.5745  & 0.6713            \\ \midrule
\multirow{5}{*}{Spanish}    & MUSE + CNN-GRU        &0.4382    &0.3354    &0.3558    &0.4203     &0.4995     &0.6364               \\
                            & Translation + BERT    &\underline{0.4598}    &\underline{0.4722}    &\underline{0.5080}    &0.4576     &\underline{0.6035}     &\underline{0.7237}               \\
                            & LASER + LR             &\textbf{0.5168}    &\textbf{0.5434}    &\textbf{0.5521}    &\textbf{0.5938}     &\textbf{0.6153}  &0.6997\\
                            & mBert                  &0.4395  &0.4285  & 0.4048 & \underline{0.4861}  &   0.5999  & \textbf{0.7329}               \\ \midrule
\multirow{5}{*}{French}     & MUSE + CNN-GRU        &\underline{0.4878}    &\underline{0.4683}    &\underline{0.5008}    &\underline{0.5222}     &0.5250     &0.5619               \\
                            & Translation + BERT    &0.4173    &0.4260    &0.4429    &0.4749     &\underline{0.6037}     &\textbf{0.6595}               \\
                            & LASER + LR             &\textbf{0.5058}    &\textbf{0.5486}    &\textbf{0.6136}    &\textbf{0.6302}     &\textbf{0.6085}     &\underline{0.6172}\\
                            & mBert                  & 0.4818 & 0.4139 & 0.4053 & 0.4355  & 0.5701  & 0.6165             \\
                            \bottomrule
\end{tabular}%
}

\end{table}

\begin{table}[!htb]
\scriptsize
\centering
\caption{Multilingual scenario: the training data is from all the languages except one and the validation and testing data is from the remaining language. The \textbf{bold} figures represent the best scores.}
\label{tab:allbutone}
\begin{tabular}{cccccccccc}
\toprule
\multirow{2}{*}{Testing Language}   & \multirow{2}{*}{Model} & \multicolumn{6}{c}{Training Size}         \\ 
                            &                        & Zero shot& 16 & 32 & 64 & 128 & 256 & Full D \\ \midrule
\multirow{2}{*}{Arabic}     
                            & LASER + LR             &0.4645    &\textbf{0.4651}    &0.4664    &0.4704     &0.4784     &0.4930   &0.6751       \\
                            & mBert                  &\textbf{0.6442}    &0.4535     &\textbf{0.4738}     &\textbf{0.5302}   &\textbf{0.7331}     &\textbf{0.7707}   &\textbf{0.8365}        \\ \midrule
\multirow{2}{*}{English}    
                            & LASER + LR             &\textbf{0.6050}    &\textbf{0.6051}    & \textbf{0.6052}     &\textbf{0.6053}     &\textbf{0.6054}     &0.6060    &0.6808       \\
                            & mBert                  &0.4971    &0.4750     &0.4670     &0.5044     &0.5242     &\textbf{0.6091}     &\textbf{0.7374}               \\ \midrule
\multirow{2}{*}{German}     
                            & LASER + LR             &0.4695    &0.4661    &0.4727    &0.4729     &\textbf{0.4740}     &0.4784   &0.5622        \\
                            & mBert                  &\textbf{0.5437}    &\textbf{0.5146}    &\textbf{0.4927}     &\textbf{0.4733}     &0.4718    & \textbf{0.4786}  &\textbf{0.6651}             \\ \midrule
\multirow{2}{*}{Indonesian}             
                            & LASER + LR             &\textbf{0.6263}    &\textbf{0.6251}    &\textbf{0.6252}      &\textbf{0.6241}     &\textbf{0.6182}     &0.6151     &0.5977        \\
                            & mBert                  &0.5113    &0.5186    &0.5049      &0.4871     &0.5864     &\textbf{0.6318}     &\textbf{0.8044}            \\ \midrule
\multirow{2}{*}{Italian}                 
                            & LASER + LR             &\textbf{0.6861}    &\textbf{0.6857}    &\textbf{0.6855}    &\textbf{0.6855}     &\textbf{0.6860}     &0.6867   & 0.7071       \\
                            & mBert                  &0.5335    &0.5318   &0.5444   &0.6696     &0.6704         &\textbf{0.7189}  & \textbf{0.8147}             \\ \midrule
\multirow{2}{*}{Polish}                   
                            & LASER + LR             &\textbf{0.5912}    &\textbf{0.5926}    &\textbf{0.5931}    &\textbf{0.5935}     &\textbf{0.5901}     &\textbf{0.5829} &  0.5672     \\
                            & mBert                  &0.0725    &0.4961     &0.5049     &0.4841     &0.4842     &0.4842   &\textbf{0.6670}          \\ \midrule
\multirow{2}{*}{Portuguese}               
                            & LASER + LR             &\textbf{0.6567}    &\textbf{0.6565}    &\textbf{0.6566}    &\textbf{0.6563}     &\textbf{0.6565}     &\textbf{0.6573}   & \textbf{0.6755}    \\
                            & mBert                  &0.5995    &0.5526   &0.5694     &0.5961     &0.6148   &0.6294     &0.6660            \\ \midrule
\multirow{2}{*}{Spanish}                 
                            & LASER + LR             &\textbf{0.5408}    &\textbf{0.5415}    &\textbf{0.5417}    &\textbf{0.5406}     &0.5434     &0.5437   &0.5708        \\
                            & mBert                  &0.2677    &0.4464   &0.4751    &0.5126     &\textbf{0.6080}     &\textbf{0.6302}     &\textbf{0.7383}               \\ \midrule
\multirow{2}{*}{French}                  
                            & LASER + LR             &0.4228    &0.4180    &0.4171    &0.4180     &0.4181     &0.4198   &0.4684            \\
                            & mBert                  &\textbf{0.5487}    &\textbf{0.5310}     &\textbf{0.5138}     &\textbf{0.5698}     &\textbf{0.5849}     &\textbf{0.5948}  & \textbf{0.5968}             \\
                            \bottomrule
\end{tabular}

\end{table}

\subsection{Multilingual scenario}
In this setting, we will use the dataset from all the languages expect one $(N-1)$, and use the validation and test set of the remaining language. This scenario represents when one wishes to employ the existing hate speech dataset to build a classifier for a new language. We have considered \textbf{LASER + LR} and \textbf{mBERT} that are most relevant for this analysis.
In the \textbf{LASER + LR} model, we take the LASER embeddings from the $(N-1)$ languages and add to this the target language data points in incremental steps of $16,32,64,128$ and 256. The logistic regression model is trained on the combined data, and we test it on the held out test set of the target language. 

For using the multilingual setting in \textbf{mBERT} we adopt a two-step fine-tuning method. For a language \textit{L}, we use the dataset for $N-1$ languages (except the $L^\textrm{th}$ language) to train the \textbf{mBERT} model. On this trained \textbf{mBERT} model, we perform a second stage of fine-tuning using the training data of the target language in incremental steps of $16,32,64,128,256$. The model was then evaluated on the test set of the $L^\textrm{th}$ language.

We also test the models for zero shot performance. In this case, the model is not provided any data of the target language. So, the model is trained on the $(N-1)$ languages and directly tested on the $N^\textrm{th}$ language test set. This would be the case in which we would like to directly deploy a hate speech classifier for a language which does not have any training data.

\noindent\textbf{Observations:} Table~\ref{tab:allbutone} reports the results of the multilingual scenario. Similar to the monolingual scenario, we observe that with increasing training data, the classifier performance increases in general. 

This is especially true in low resource settings of the target languages such as English, Indonesian, Italian, Polish, Portuguese.

In case of zero shot evaluation, we observe that \textbf{mBERT} performs better than \textbf{LASER + LR} in three languages (Arabic, German, and French). \textbf{LASER + LR} perform better on the remaining six languages with the results in Italian and Portuguese being pretty good. In case of Portuguese, zero shot \textbf{Laser + LR} (without any Portuguese training data) obtains an F-score of 0.6567, close to the best result of 0.6941 (using full Portuguese training data).

For the languages such as Arabic, German, and French, \textbf{mBERT} seems to be performing better than \textbf{LASER + LR} is almost all the cases (low resource and Full D). \textbf{LASER + LR}, on the other hand, is able to perform well for Portuguese language in all the cases. For the rest of the five languages, we observe that \textbf{LASER + LR} is performing better in low resource settings, but on using the full training data of the target language, \textbf{mBERT} performs better.

\subsection{Possible recipes across languages}
As we have used the same test set for both the scenarios, we can easily compare the results to access which is better. Using the results from monolingual and multilingual scenario, we can decide the best kind of models to use based on the availability of the data. The possible recipes are presented as a catalogue in Table~\ref{tab:recipe}. Overall we observe that \textbf{LASER + LR} model works better for low resource settings while BERT based models work well for high resource settings. This possibly indicates that BERT based models, in general can work well when there is larger data available thus allowing for a more accurate fine-tuning. We believe that this catalogue is one of the most important contributions of our work which can be readily referred to by future researchers working to advance the state-of-the-art in multilingual hate speech detection.

\begin{table}[!htb]
\centering
\caption{The table describes the best model to use in low and high resource scenario. In general, LASER + LR performs well in low resource setting and BERT based models are better in high resource settings}
\label{tab:recipe}
\begin{tabular}{c cc}
\hline
\textbf{Language} & \textbf{Low resource}  & \textbf{High resource} \\ \hline
Arabic            & Monolingual, LASER + LR  & Multilingual, mBERT    \\
English           & Multilingual, LASER + LR & Multilingual, mBERT    \\
German            & Monolingual, LASER + LR  & Translation + BERT     \\
Indonesian        &Multilingual, LASER + LR  & Monolingual, mBERT     \\
Italian           & Multilingual, LASER + LR  & Monolingual, mBERT     \\
Polish            &  Multilingual, LASER + LR & Translation + BERT     \\
Portuguese         & Multilingual, LASER + LR & Monolingual, LASER+LR  \\
Spanish           & Monolingual, LASER + LR  & Multilingual, mBERT    \\
French            & Monolingual, LASER + LR  & Translation + BERT     \\ \hline
\end{tabular}
\end{table}

\section{Discussion and Error Analysis}\label{sec:discussion}

\subsection{Interpretability}

 In order to compare the interpretability of \textbf{mBERT} and \textbf{LASER + LR}, we use LIME~\cite{ribeiro2016should} to calculate the average importance given to words by a particular model. We compute the top 5 most predictive words and their attention for each sentence in the test set. The total score for each word is calculated by summing up all the attentions for each of the sentences where the word occurs in the top 5 LIME features. The average predictive score for each word is calculated by dividing this total score by the occurrence count of each word. In Table~\ref{Tab:interpretability} we note the top 5 words having the highest attention scores and compare them qualitatively across models. 

\begin{table}[!ht]
\caption{Interpretations of the model outcomes.}
\label{Tab:interpretability}
\resizebox{\textwidth}{!}{%
\begin{tabular}{cc|cc}
\hline

 \multicolumn{2}{c}{\textbf{German}} & \multicolumn{2}{c}{\textbf{Indonesian}} \\
 \underline{mBERT}        & \underline{LASER + LR}       & \underline{mBERT}          & \underline{LASER + LR} \\
spendieren (spend) & fotzen (pu**ies) & loo (loo)& NAJIS (unclean) \\
drogen (drugs) & Trottel (fool) & rusak (broken)& bajingan (son of a bi**h)  \\
schœn (beautiful) & abschaum (scum) & makhluk (creature)& MAMPUS (dead)   \\
kastrieren (castrate) & WICHSER (w**ker) & pengkhianatan (betrayal) & Idiot (idiot)   \\
einsetzen (deploy) & Scheissen (shit) & celeng (wild boar)& F**kYou (f**k you)    \\ \hline
 
 \multicolumn{2}{c}{\textbf{Italian}} & \multicolumn{2}{c}{\textbf{Polish}} \\
 \underline{mBERT}        & \underline{LASER + LR}       & \underline{mBERT}          & \underline{LASER + LR} \\
 innervosirmi (get nervous) & Schifo (schifo) & stanowisk (posts) &  pieprzysz (f**k) \\
  vomitata (vomited) & demoliscile (demoliscile) & pomysł (idea) &  gówno (shit) \\
  cascarci (fall for) & disonesti (dishonest) & powiedzieli (they said) &  idiota (idiot) \\
  italioti (italioti) & massacrale (massacrale) & cwelica (cwelica) &  Idiotów (idiots) \\
  annegano (drown) & schifoso (lousy)   & obrazka (picture) &  świry (suck) \\   \hline

  \multicolumn{2}{c}{\textbf{Portuguese}} & \multicolumn{2}{c}{\textbf{Spanish}} \\
 \underline{mBERT}        & \underline{LASER + LR}       & \underline{mBERT}          & \underline{LASER + LR} \\
  fuder (f**k)   & FOFURA (cuteness) &  Hxrry\_again (hxrry\_again) & piratas (pirates)	\\
 heterofobicos (heterophobic)    & tretas (fights) &    majisimos  (majestic) & MARICA (sissy)    	\\
 vagabunda (slut)    & porcaria (filth) &   mate (mate) &   perseguidos (persecuted)	\\
 cracuda (crunchy)   & foda (f**k) &    publicidad (advertising) &  pegaso6038 (pegasus6038)	\\
 femimimismo (feminism)  & heterofobicos (heterophobic) &   sevilla (seville) & Putas (wh**es)	\\ \hline
 
   \multicolumn{2}{c}{\textbf{French}} &\\
    \underline{mBERT}        & \underline{LASER + LR}       &    &  \\
mongol (mongolian)	&   jérusalem  (jerusalem)    	 &    & \\
medelin (medelin)	&   ptdrrrrrrrrrrr (ptdrrrrrrrrrrr)	 &    & \\
arabe (arab)	&   negrophobe (ne*rophobe)	 &    & \\
barges (barges)	&   juifs (jews)	 &    & \\
marocains (moroccons)	&   bf (bf)	 &    & \\
  
\cline{1-2}
\end{tabular}}%
\end{table}

While comparing the models' interpretability in Table \ref{Tab:interpretability}, we see that \textbf{LASER + LR} focuses more on the hateful keywords compared to \textbf{mBERT}, i.e., words like `pigs' etc. \textbf{mBERT} seems to search for some context of the hate keywords as shown in Table \ref{tab:lime_examples}. Models dependent on the keywords can be useful when we are in a highly toxic environment such as GAB\footnote{\url{https://en.wikipedia.org/wiki/Gab_(social_network)}} since most of the derogatory keywords typically occur very close or at least simultaneously along with the hate target,for e.g., the first case in Table~\ref{tab:examples}. \if{0}It may be noted that we do not replace our model with a keyword based model \punyajoy{Why ?}.\fi In sites which are less toxic like Twitter, complex methods giving attention to the context like \textbf{mBERT} might be more helpful,for e.g., the third case in Table~\ref{tab:examples}. 

\begin{table}[]
\centering
\caption{\label{tab:lime_examples}Examples showing word with the highest predictive word for both \textcolor{red}{\textbf{mBERT}} and \textcolor{blue}{\textit{LASER + LR}}.}
\begin{tabularx}{\textwidth}{l}
\hline
\textbf{sentences with hate label}\\
\hline
\begin{tabular}[c]{@{}l@{}}das \textcolor{blue}{\textit{pack}} muss tag und nacht \textcolor{red}{\textbf{gejagt}}  werden,ehe sie es mit den deutschen machen !!\\ (\textbf{Translated :-} the \textcolor{blue}{\textit{pack}} must be \textcolor{red}{\textbf{hunted}} day and night before they do it with \\ the Germans !!)\end{tabular} \\
\hline
\begin{tabular}[c]{@{}l@{}}absolument ! il faut l'arraisonner en mer par la marin nationale arrêter tous les \\ occupants \textcolor{red}{\textbf{expulser}} les \textcolor{blue}{\textit{migrant}}… @url (\textbf{Translated :-} absolutely! it must be \\ boarded at  sea by the navy national arrest all occupants \textcolor{red}{\textbf{expel}} \textcolor{blue}{\textit{migrants}}… @url)
 \end{tabular} \\
\hline
\end{tabularx}
\end{table}

\subsection{Error Analysis}

In order to delve further into the models, we conduct an error analysis\footnote{Note that we rely on translation for interpretations of the errors and the translation itself might also have some error.} on both the \textbf{mBERT} and \textbf{LASER + LR} models using a sample of posts where the output was wrongly classified from the test set.We analyze the common errors and categorize them into the following four types:

\begin{compactenum}
    \item \textbf{Wrong classification due to annotation's dilemma (AD):} These error cases occur due to ambiguous instances where according to us the model predicts correctly but the annotators have labelled it wrong.  
    \item \textbf{Wrong classification due to confounding factors (CF):} These error cases are caused when the model predictions rely on some irrelevant features like normalized form of mentions (\textit{@user}) and links (\textit{URL}) in the text.  
    \item \textbf{Wrong classification due to hidden context (HC):} These error cases are caused when the model fails to capture the context of the post.
    \item \textbf{Wrong classification due to abusive words (AW):} These error cases are caused by over-dependence of the model on the abusive words.
\end{compactenum}

Table~\ref{tab:mBERT_error} shows the errors of the \textbf{mBERT} and \textbf{LASER + LR} models. For \textbf{mBERT}, the first example has no specific indication of being a hate speech and is considered an error on the part of annotators. In the second example the author of the post actually wants the reader to not use the abusive terms, i.e., sl*t and wh*re (\textit{found using LIME}) but the model picks them as indicators of hate speech. The third example has mentioned the term ``parasite" as a derogatory remark to refugees and the model did not understand it.

For the \textbf{LASER + LR} model, the first example is an error on the part of the annotators. In the second case the model captures the word ``USER" (\textit{found using LIME}), a confounding factor which affects the models' prediction. For the third case, the author says (s)he will leave before homosexuality gets normalized which shows his/her hatred toward the LGBT community but the model is unable to capture this. In the last case the model predicts hate speech based on the word ``retarded" (\textit{found using LIME)} which should not be the case.

\begin{table}[!htb]
\centering
\caption{Various types of errors \textbf{(E)} for the models \textbf{(M)} : mBERT and LASER + LR. The ground truth \textbf{(GT)} and prediction \textbf{(P)} consist of 0 (Non-Hate)/1 (Hate) label.} 
\label{tab:mBERT_error}
\begin{tabular}{ M{0.2in} | p{0.8\textwidth} | p{0.05\textwidth} p{0.03\textwidth} p{0.04\textwidth} }
\textbf{M}&\textbf{Sentences} & \textbf{GT} & \textbf{P} & \textbf{E}\\
\hline
& \textbf{Arabic Translation:} He and his father, and Abu Alto and Abu Israel, are doomed to go to Israel to blind, insolent Syrian opponents, and to betray that I have not seen and my eyes have seen. & 1 & 0 & AD \\ \cline{2-5}
\multirow{3}{*}{\rotatebox[origin=c]{90}{mBERT}} &``If you have tries to get w/a girl you are not allowed to call her demeaning names like ``slut whore etc" sorry bout yall"& 0 & 1 & AW\\
\cline{2-5}
&``Könnten wir Schmarotzer und Kriminelle  loswerden würde die Asylanten-Schwemme  auf beherrschbare Zahlen runtergehen." \newline \textbf{Translation:} If we could get rid of parasites and criminals, the  asylum seeker  flood would drop to manageable numbers.
& 1 & 0 & HC\\
\hline \hline
 &``Die hat jede Art von Realität verloren und braucht dringend Hilfe am besten ne Einweisung in die Geschlossene für immer und Ewig und ihr Gefolge gleich mit" \newline \textbf{Translation:} She has lost all kind of reality and urgently needs help, best a briefing in the closed forever and ever and her followers at the same time & 1 & 0 & AD\\
\cline{2-5}
&``USER USER Gw mah tetep anti cina... gara gara gw ngga bisa sipit dan putih kayak mereka...wkwkwk " \newline
\textbf{Translation:} USER USER I am still anti-Chinese ... because I can't be narrow  and white like them ... hahaha
& 0 & 1 & CF\\
\cline{2-5}
\multirow{4}{*}{\rotatebox[origin=B]{90}{LASER + LR}}&``RT @mundodrogado: Antes o homossexualismo era proibido.Depois passou a ser tolerado.Hoje é normal.  Eu vou embora antes que vire obrigatór \newline
\textbf{Translation:} RT @mundodrogado: Before homosexuality was forbidden. Then it  became tolerated. Today it's normal. I'm leaving before it becomes mandatory ...& 1 & 0 & HC  \\
\cline{2-5}
&this movie is actually good cuz its so retarded
& 0 & 1 & AW\\
\hline

\end{tabular}
\end{table}

\section{Conclusion}
In this paper, we perform the first large scale analysis of multilingual hate speech. Using 16 datasets from 9 languages, we use deep learning models to develop classifiers for multilingual hate speech classification. We perform many experiments under various conditions -- low and high resource, monolingual and multilingual settings -- for a variety of languages. Overall we see that for low resource, LASER + LR is more effective while for high resource BERT models are more effective. We finally suggest a catalogue which we believe will be beneficial for future research in multilingual hate speech detection.

%
%
\bibliographystyle{splncs04}
\bibliography{main}
%





\end{document}